\documentclass[sigconf,natbib=false]{acmart}
\AtBeginDocument{%
  \providecommand\BibTeX{{%
    \normalfont B\kern-0.5em{\scshape i\kern-0.25em b}\kern-0.8em\TeX}}}

\copyrightyear{2024} 
\acmYear{2024} 
\setcopyright{rightsretained} 
\acmConference[ASE '24]{39th IEEE/ACM International Conference on Automated Software Engineering }{October 27-November 1, 2024}{Sacramento, CA, USA}
\acmBooktitle{39th IEEE/ACM International Conference on Automated Software Engineering (ASE '24), October 27-November 1, 2024, Sacramento, CA, USA}
\acmDOI{10.1145/3691620.3695021}
\acmISBN{979-8-4007-1248-7/24/10}

\RequirePackage[
  datamodel=acmdatamodel,
  style=acmnumeric,
  ]{biblatex}
  
\usepackage{multirow}
\usepackage[most]{tcolorbox}
\usepackage{pifont}
\usepackage{array} 
\usepackage{helvet} 
\usepackage{xcolor}
\usepackage{hyperref}
\usepackage{todonotes}
\usepackage{fancyvrb}
\usepackage{xurl}
\usepackage{enumitem}
\usepackage{soul} 
\usepackage{graphicx}

\usepackage{xcolor}
\usepackage{parskip}
\definecolor{darkgreen}{RGB}{0,160,0}

\newcommand{\revisionmarker}[2]{%
  \noindent\textcolor{darkgreen}{\textbf{#1 #2}}%
  \par\noindent%
}

\newcommand{\revisionstart}[2]{\revisionmarker{=====Begin #1}{: #2=====}}
\newcommand{\revisionend}[2]
{\revisionmarker{=====End #1}{: #2=====}}

\usepackage{listings}
\usepackage{mdframed}

\hypersetup{
    colorlinks=true,
    linkcolor=blue, 
    filecolor=magenta,      
    urlcolor=blue, 
    citecolor=blue, 
}

\usepackage{booktabs}
\usepackage[font=small,skip=0pt]{caption}

\usepackage{titlesec}
\titlespacing*{\paragraph}{0pt}{0.5em}{0.5em} %

\newcommand{\xmark}{\textcolor{red}{\ding{55}}}    

\newcommand{\edits}[1]{\textcolor{darkgreen}{#1}} 

\newtcolorbox{summarybox}[1]{
  colback=yellow!10, 
  colframe=blue!50!black, 
  title=#1, 
  fonttitle=\bfseries,
  sharp corners,
  rounded corners=southeast,
  arc is angular,
  arc=3mm,
  boxrule=0.4pt,
  left=2mm,
  right=2mm,
  top=1mm,
  bottom=1mm,
  before skip=2mm,
  after skip=3mm
}

\begin{document}

\title[Promise and Peril of Collaborative Code Generation Models]{Promise and Peril of Collaborative Code Generation Models: Balancing Effectiveness and Memorization}


\author{Zhi Chen, Lingxiao Jiang}
\affiliation{%
  \institution{Centre for Research on Intelligent Software Engineering\\
  School of Computing and Information Systems\\
  Singapore Management University}
  \vspace{-2mm}  
  \country{Singapore}
}

\email{{zhi.chen.2023, lxjiang}@smu.edu.sg}


\begin{abstract}
In the rapidly evolving field of machine learning, training models with datasets from various locations and organizations presents significant challenges due to privacy and legal concerns. The exploration of effective collaborative training settings, which are capable of leveraging valuable knowledge from distributed and isolated datasets, is increasingly crucial.This study investigates key factors that impact the effectiveness of collaborative training methods in code next-token prediction, as well as the correctness and utility of the generated code, showing the promise of such methods. Additionally, we evaluate the memorization of different participant training data across various collaborative training settings, including centralized, federated, and incremental training, showing their potential risks in leaking data.

Our findings indicate that the size and diversity of code datasets are pivotal factors influencing the success of collaborative trained code models. We demonstrate that federated learning achieves competitive performance compared to centralized training while offering better data protection, as evidenced by lower memorization ratios in the generated code. However, federated learning can still produce verbatim code snippets from hidden training data, potentially violating data privacy or copyright. Our study further explores the patterns of effectiveness and memorization in incremental learning, emphasizing the importance of the sequence in which individual participant datasets are introduced. Also, we identify the memorization phenomenon of cross-organizational clones as a prevalent challenge in both centralized and federated learning scenarios. Our findings highlight the persistent risk of data leakage during inference, even when training data remains unseen. We conclude with strategic recommendations for practitioners and researchers to optimize the use of multisource datasets, thereby propelling the cross-organizational collaboration forward.


\end{abstract}




\maketitle
\section{Introduction}
\begin{figure*}[!htb]
    \centering
    \includegraphics[width=\textwidth]{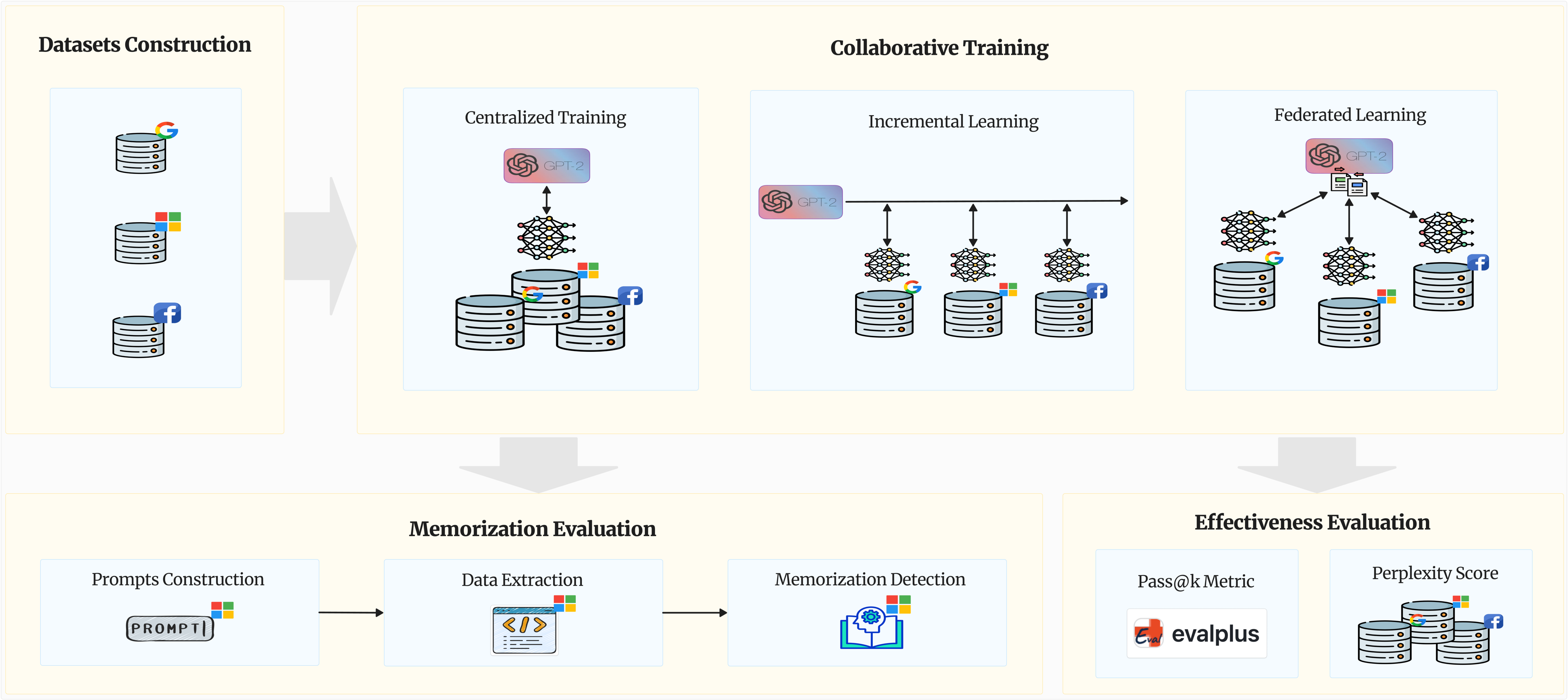}
    \begin{flushleft}
    \colorbox{gray!10}{
        \parbox{\dimexpr\textwidth-4\fboxsep}{
        \textbf{Description:} This figure illustrates the overall workflow of our study, which includes constructing cross-organizational datasets, collaboratively training code models using various methods (centralized, federated, and incremental learning), and evaluating the models. The evaluation of effectiveness encompasses next token prediction accuracy, as well as the correctness and utility of the generated code. The assessment of memorization in the participants' training data involves prompt construction, data extraction, and memorization detection. Specific tools, benchmarks, and metrics are employed for detailed analysis.
        }
    }
    \end{flushleft}
    \caption{Overview}
    \label{fig:overview}
\end{figure*}

Large language models for code~\cite{ jiang2023impact, nijkamp2022codegen,zhuo2023popquizpretrainedcode} automatically generate code snippets, functions, or entire programs based on given inputs, significantly enhancing developer productivity and aiding in software development~\cite{du2024evaluating,xia2023automated,zhang2023multilingual}. Effective training of code generation models requires large and diverse source code datasets. However, reliance on open-source repositories is becoming increasingly unsustainable. For instance, StarCoder2 \cite{li2023starcoder,lozhkov2024starcoder} has been trained on a massive dataset aggregated from various platforms like GitHub and Kaggle, demonstrating that current models have nearly exhausted the available open-source training data. Moreover, open-source datasets pose significant risks, including the presence of vulnerable or malicious code and legal concerns related to the commercial use of copyleft-licensed code~\cite{sun2022coprotector}. Studies on GitHub Copilot have shown that models can inherit vulnerabilities from unvetted code~\cite{mcmahan2017communication}. Given these concerns, there is a growing need to use collaborative approaches to explore the untapped value of proprietary (closed-source) code datasets from different organizations~\cite{hoang2024collaborative,sim2024incentives}.

Several collaborative training methods are available, but privacy concerns remain a significant obstacle. Traditional centralized training is effective when data can be aggregated~\cite{truong2021privacy}, but due to privacy concerns, such as sensitive information and legal constraints on data sharing, it becomes impractical~\cite{henze2016moving,liu2021fate}. These challenges necessitate the exploration of privacy-preserving collaborative learning methods such as federated learning~\cite{lo2021systematic} and incremental learning~\cite{gepperth2016incremental}. Federated learning allows for collaborative training without centralizing data, enabling participants to maintain control over their private datasets~\cite{hard2018federated,li2020review,yang2019federated}. Incremental learning, which updates models gradually with new data, offers a promising solution for dynamic environments where data continuously evolves~\cite{perez2010incremental}.

However, even though certain methods can protect privacy by ensuring data remains unseen during training, studies on data extraction attacks reveal that models can still leak training data due to memorization~\cite{carlini2021extracting,Al-Kaswan2024memorization,yang2024unveiling}. This is a significant concern that may discourage organizations from providing their private data for collaborative training~\cite{banabilah2022federated,zhang2022federated}. Moreover, preprocessing cross-organizational datasets poses a substantial challenge due to the unseen nature of other participants' data. One issue is the presence of cross-organizational clones, since code clones not only waste computing resources on duplicates but also increase the likelihood of these clones being memorized~\cite{yang2024unveiling}. Ideally, the model should learn to generalize from the training data and develop the capability to generate new code, rather than reproducing the training data verbatim due to memorization. This is crucial because using verbatim generated code that is the same as some copyrighted code can lead to legal issues, as demonstrated by Oracle's lawsuit against Google. In this case, Google developed Android without a Java license and copied its APIs, resulting in a copyright infringement case over "nine lines of code".\footnote{Google LLC v. Oracle America, Inc. \url{https://en.wikipedia.org/wiki/Google_LLC_v._Oracle_America,_Inc.}}

\paragraph{\textbf{Key Research Question.}} 
The main objective of this paper is to better understand the promise and peril of collaborative training in the context of code generation task using several cross-organizational code datasets. This research aims to investigate a key question:

\begin{quote}
\textit{How do the effectiveness and memorization patterns vary in code models trained under different collaborative training settings?}
\end{quote}

Our investigation underscores the critical impact of dataset size and diversity on the effectiveness of collaborative trained code models, where effectiveness is measured as the model's next token prediction ability and the correctness and utility of its generated code.
We found that federated learning approaches yield results comparable to centralized training while maintaining data confidentiality during training and showing lower memorization rates during inference. Centralized training, however, tends to exhibit increased memorization, particularly with duplicate-heavy datasets. Both centralized and federated models showed higher memorization of cross-organizational clones than incremental models. Additionally, the effectiveness and memorization tendencies of incremental learning heavily depend on the order of participant datasets introduction. Crucially, our findings highlight the ongoing threat of data exposure during the inference stage, even without direct observation of the training data.




\paragraph{\textbf{Main Contributions.}}
\begin{itemize}[leftmargin=2em]
    \item We have conducted a comprehensive analysis of various collaborative training setups, assessing the impact of dataset size, diversity, and data presentation sequence on the effectiveness of these methods for code generation.
    \item To the best of our knowledge, we are the first to systematically examine the phenomenon of training data memorization in various centralized, federated, and incremental learning settings, identifying the associated risks of training data leakage.
    \item Our findings provide actionable insights and recommendations for industry professionals and academic researchers, aiming to facilitate collaborative training practices, maximize the potential of extensive, multisource code repositories, and minimize the risk of code leakage. These insights ultimately urge the enhancement of privacy- and copyright-preserving capabilities of large code models while propelling cross-organizational collaboration forward.
\end{itemize}


\paragraph{\textbf{Paper Structure.}}
Section \ref{sec:methodology} details the methodology of our study.
Section \ref{sec:dataset} describes the datasets we collected.
Section 4 describes the experimental setup.
Section 5 presents our evaluation results, analyzing model effectiveness and memorization patterns. 
Section 6 discusses the findings and threats to validity.
Section 7 reviews related work.
Section 8 concludes with key findings.

\section{Methodology}
\label{sec:methodology}

In this section, we present our specific research questions (Section \ref{sec:RQs}) and the workflow and tools employed to answer the questions.
The workflow of our study is illustrated in Figure \ref{fig:overview}.
It begins with the explanation of our dataset construction method (Section \ref{sec:data-construction}), followed by a description of the collaborative training methods we used to train models (Section \ref{sec:colmethods}). Subsequently, we outline the method and metrics used to evaluate the effectiveness of the trained models (Section \ref{sec:effect-eval}). Finally, we detail the training data extraction techniques and memorization evaluation methods employed to assess the extent of data memorization (Section \ref{sec:mem-eval}).

\subsection{Research Questions}
\label{sec:RQs}

We aim to investigate the promise and peril of collaborative training in the following research questions.

\paragraph{\textbf{RQ1.} What factors most significantly impact the effectiveness of collaborative training methods for code generation models?}
\paragraph{Motivation.} To enhance the practical utility of code generation models trained on diverse datasets from multiple organizations, it is essential to understand how different factors influence collaborative training methods. By exploring how the size and diversity of datasets, as well as the sequence of data presentation, impact the performance of these models, we can derive valuable insights. This research seeks to identify these factors to inform the development of collaborative training strategies—such as centralized training, federated learning, and incremental learning—that optimize model effectiveness and support their application in real-world scenarios.

\paragraph{\textbf{RQ2.} To what extent is data from different participants memorized in various collaborative training settings?}

\paragraph{Motivation:} Privacy concerns regarding the potential leakage of sensitive training data pose a significant barrier to organizational participation in collaborative training. Even with techniques like federated learning and incremental learning, which ensure that training data remains unseen during the training process, there remains a risk of data leakage through memorization during inference. Understanding how data from different participants is memorized and uncovering the memorization patterns can provide insights for improving collaborative training methods to mitigate memorization risks and enhance privacy or copyright preservation, thereby encouraging more organizations to engage in collaborative training and increasing the utility of valuable untapped proprietary datasets.

\paragraph{\textbf{RQ3.} How are cross-organizational code clones memorized in collaborative models?}

\paragraph{Motivation:} 
Collaborative training scenarios present unique challenges, particularly concerning cross-organizational code clones. While centralized training can efficiently remove these clones, federated learning and incremental learning prevent participants from performing cross-dataset checking and filtering. This limitation can lead to the persistence of cross-organizational clones. A higher occurrence of code clone snippets can increase the risk of unintentional verbatim code exposure~\cite{yang2024unveiling}. For instance, clones might include licensed code reused properly within organizations, but if a model reproduces this code verbatim due to memorization, users might unknowingly misuse these clones, potentially violating licensing regulations. Additionally, the quality of the generated code could be compromised if these clones contain vulnerabilities. This RQ is to evaluate how these clones are memorized in code models trained under different collaborative settings, providing insights into memorization patterns and highlighting the need for specialized dataset preprocessing in collaborative training scenarios.

\subsection{Dataset Construction Method}
\label{sec:data-construction}

Our investigation on collaborative training naturally needs datasets from different participants or organizations. Although we cannot use real-world proprietary codebases, we can construct separate datasets from open-source code repositories to simulate multisource datasets for our evaluation.

\paragraph{\textbf{Cross-Organizational Datasets Construction Approach.}}
Due to the difficulty of obtaining proprietary code datasets from industry sources for collaborative training, our methodology involves collecting cross-organizational datasets from GitHub repositories while adhering to the following principles:
\begin{itemize}[leftmargin=2em,nosep]
    \item Ensuring that the code in one dataset comes from a single organization while the code in different datasets comes from different organizations, simulating scenarios where each participant in a collaborative training setting has their own private codebase.
    \item Limiting the datasets to a single programming language to facilitate more consistent evaluation of effectiveness and memorization issues in the trained models.
\end{itemize}

Based on these principles, our methodology involves curating Python code files from the open-source repositories of three prominent tech organizations hosted on GitHub: Facebook (F)\footnote{Facebook is now Meta. Although the company has undergone a rebranding, many repositories on GitHub continue to use the name Facebook. Therefore, for consistency and clarity within this context, we will refer to the company as Facebook (F).}, Microsoft (M), and Google (G).
These Python code files are primarily developed by internal software engineers from these organizations,
enabling effective simulation of collaborative training methods in real-world scenarios.

\paragraph{\textbf{Data Collection Platform.}}
We utilize Google's BigQuery to collect our code datasets as it contains extensive GitHub data\footnote{https://cloud.google.com/blog/topics/public-datasets/github-on-bigquery-analyze-all-the-open-source-code}. This GitHub data on BigQuery is updated weekly and can be accessed through efficient SQL queries, ensuring timely access to the latest GitHub data.
Section \ref{sec:dataset} gives more details about our collected codebases.

\subsection{Collaborative Training Methods}
\label{sec:colmethods}

There are different ways to perform collaborative training using datasets from different participants or organizations.
We summarize three common methods in Table \ref{table:collaborative_training_methods}:
traditional \textit{centralized training}, ideal for mutually trusting participants who combine datasets on a centralized location; and \textit{federated learning} and \textit{incremental learning}, able to train models in a decentralized manner. The latter two methods prevent dataset centralization, ensuring that the training data remains unseen during the training process, and consequently, to some extent, safeguard the privacy of the source data.

\begin{table}[t]
\caption{Collaborative Training Methods}
\label{table:collaborative_training_methods}
\centering
\begin{tabular}{lcc}
\toprule
\text{Method} & \text{Decentralized?} & \text{Synchronous?} \\ \hline
Centralized Training & \textcolor{red}{\xmark} & \textcolor[rgb]{0.0, 0.5, 0.0}{\checkmark} \\
Federated Learning & \textcolor[rgb]{0.0, 0.5, 0.0}{\checkmark} & \textcolor[rgb]{0.0, 0.5, 0.0}{\checkmark} \\
Incremental Learning & \textcolor[rgb]{0.0, 0.5, 0.0}{\checkmark} & \textcolor{red}{\xmark} \\ \bottomrule
\end{tabular}
\end{table}

In terms of the synchronicity in the training process across different datasets, that is, whether in each training round (epoch) of a model, the data from all parties are involved in the training and contribute to the model’s update, we classify the three methods into \textit{Synchronous Collaborative Training} (e.g., centralized training and federated learning) and \textit{Asynchronous Collaborative Training} (e.g., incremental learning with sequential dataset training).

We provide a detailed explanation of the three methods using a unified representation, to better illustrate the collaborative training approaches utilized in this study.


\subsubsection{Dataset and Model Representation}

\paragraph{Datasets.}
We use $D_i$ to denote a dataset from a participant $i$. Given $n$ participants, the centralized union of all their datasets is denoted as $D_C = \cup_{i=1}^{n} D_i$.
In our study, we have three datasets $D_F$, $D_M$, $D_G$ from Facebook, Microsoft, Google, respectively. Each data point in the dataset can be a Python code file, a Python class, or a function, optionally associated with some docstrings or comments. These datasets will be used in various ways to train various models for code generation tasks in our evaluation.

\paragraph{Models.}
Our study focuses on models that are based on deep neural networks, as they have been shown to be effective for code generation tasks \cite{cert2022ijcai,du2024evaluating,lozhkov2024starcoder,liu2024your}.
We denote a model $M_i$, together with its internal weights $\Theta_i$, potential inputs $X_i$, and potential outputs $Y_i$, as $Y_i=M_i(\Theta_i, X_i)$.
There may exist ground-truth outputs $\bar{Y_i}$ for the input $X_i$, and a model trained on the ground-truth data should have adjusted its internal weights $\Theta_i$ so that the differences between $Y_i=M_i(\Theta_i, X_i)$ and $\bar{Y_i}$ are minimized.

In our study, each participant $i$ can individually train a model on its own dataset $D_i$ as usual to minimize the differences between the $M_i(\Theta_i, D_i)$ and its ground truth $\bar{D_i}$. When it comes to collaborative training, the settings need to be adjusted as follows. 

\subsubsection{Centralized Training.}
This training method is ideal when two participants share a profound mutual trust. In this approach, the participants train a common model using centralized datasets that combine information from all participants. That is, the method is to train a centralized model $M_C$ so that the differences between $Y_C=M_C(\Theta_C, D_C)$ and $\bar{D_C}$ are minimized, where $D_C=D_F \cup D_M \cup D_G$.

\subsubsection{Federated Learning.}
This is a method for multiple participants to collaboratively train one central model as well while keeping their data localized \cite{shanbhag2022exploring}. This method enhances privacy and mitigates the risks associated with data centralization ~\cite{yang2019federated}.
Its key idea is for each participant to calculate the updates needed for the central model weights using their own dataset locally and only share the weight updates with all the participants. Thus, a key component of federated learning is often the aggregation strategy used to aggregate weight updates from individual participants.  

In our study, we applied two federated learning aggregation strategies, FedAvg~\cite{mcmahan2017communication} and FedYogi~\cite{reddi2021adaptive}, to diversify our experimental settings. The FedAvg algorithm~\cite{mcmahan2017communication} simply averages the model weights updated by each participant to form the global model weights.
It is often used for cases when datasets across parties are homogeneous.
That is, FedAvg trains a model $M_{FedAvg}(\Theta_{FedAvg},X)$ where $X$ is unknown, and each participant locally trains a $M_i(\Theta_i,D_i)$, and $\Theta_{FedAvg} = \frac{1}{n} \sum_{i=1}^{n} w_i \cdot \Theta_{i}$, where $w_i$ is the weight of the $i$-th participant's contribution which is often based on the size of $D_i$.
Note that the averaging operation is often done at the end of each training round (epoch).
Also, to facilitate the averaging operation, it would be better for individual $M_i$s to have the same structure (e.g., the same numbers and positions of the weights).

The FedYogi algorithm \cite{reddi2021adaptive} is similar to FedAvg, but adapts the Yogi optimizer \cite{zaheer2018adaptive} to adjust the model weights and the model learning rates for non-IID data~\cite{zhao2018federated} across participants during training.
Thus, FedYogi is often used for cases when datasets across parties are heterogeneous.


\subsubsection{Incremental Learning.}

This method involves gradual updates to a model with new datasets~\cite{wu2019large}. It is particularly useful in situations where the data evolves over time, allowing the model to adapt to the new data without being retrained from scratch~\cite{van2022three}, suitable for not only collaborative training, but also internal training with one organization.
That is, it trains a sequence of models $[M_1(\Theta_1,D_1), M_2(\Theta_2,D_2), \cdots, M_n(\Theta_n,D_n)]$ such that $\Theta_{i+1}$ are initialized with $\Theta_i$ but updated according to $D_{i+1}$ without referring back to $D_i$, and the last model $M_n$ is often used as the final collaborative model $M_I$.
Note that, to facilitate the initialization of $\Theta_{i+1}$ from $\Theta_i$, it is often better for all models $M_i$ to use the same structure.
Also, the order of using $[D_1, D_2, \cdots]$ datasets can affect the trained models.
In our study, we sequentially train various incremental models using our datasets in different orders. We use the order of the datasets used to train a model as the name of the model.
For example, we use $M_{F2M2G}$ to denote the model that is incrementally trained from the Facebook ($D_F$), Microsoft ($D_M$), and Google ($D_G$) codebases in that order. 

\subsection{Effectiveness Evaluation Method}
\label{sec:effect-eval}

This paper focuses on code generation tasks using collaborative models, which involves creating code snippets from prompts or specifications to enhance software development productivity. To evaluate the effectiveness of code generation models, we selected two primary metrics: perplexity and pass@k. These metrics provide a balanced assessment of the model's predictive capabilities and practical utility, making them the most suitable choice for RQ1 in our study.


\paragraph{\textbf{Perplexity: Evaluating Next-Token Prediction Ability.}}
Perplexity measures the model's ability to predict subsequent tokens, ensuring syntactic correctness. Lower perplexity values correspond to improved predictive performance~\cite{iyer1997analyzing}.

\paragraph{\textbf{Pass@k: Evaluating Code Correctness and Utility.}}
Pass@k for a model is defined as the probability that at least one of the top-k code samples generated by the model for a query problem passes the unit tests defined for the problem.
Higher pass@k values indicate better performance in providing relevant and accurate code solutions~\cite{chen2021evaluating}.
For each trained model, this measurement is calculated using the \textit{EvalPlus}~\cite{liu2024your} benchmark, which builds upon the \textit{HumanEval}~\cite{chen2021evaluating} benchmark. \textit{EvalPlus} enhances the scope and robustness of \textit{HumanEval} by incorporating a more diverse set of real-world coding problems.


\subsection{Memorization Evaluation Methods}
\label{sec:data-extract}
\label{sec:mem-eval}

As our research goal is to investigate the memorization of each participant's training data,
we adapt the data extraction strategies used by Al-Kaswan et al.~\cite{Al-Kaswan2024memorization}, which formulates a targeted data extraction security game to extract data from models. In the targeted attack scenario, the adversary is provided with a prefix and is tasked with recovering the suffix associated with the prefix from the training data.
Targeted attacks are more critical for security because they allow the extraction of specific information, such as sensitive configuration, personal identifiers, or proprietary algorithms \cite{10189147,liao2021generating}.

\paragraph{\textbf{Prompt Construction for Data Extraction}}
\label{sec:prompt-construction}

Different from the setting in \cite{Al-Kaswan2024memorization}, our training data is available, which allows us to construct prefix prompts directly from each organization's dataset instead of from an identified extractable dataset~\cite{Al-Kaswan2024memorization}.
For constructing prompts, we choose to use function signatures with docstrings as our "prefix" prompt. This format better reflects real-world scenarios where an adversary has access to an API's function signature and functionality description document and aims to extract the function's coding details in the function body.

Specifically, we use static analysis to parse the source code from the training data into abstract syntax trees (ASTs) to extract functions. Subsequently, two filtering conditions are applied to construct prefix prompts: each function must have a corresponding docstring, and the combined length of the tokenized function signature and docstring must not exceed 512 tokens.\footnote{GPT2 can only handle a total token length of 1024 (including input and newly generated tokens), so we set the maximum length of the input and the maximum length of the newly generated tokens to 512, respectively.}
Listing \ref{lst:prompt_format} provides a concrete example of the function prompt utilized in our evaluation.

\begin{center}
\begin{minipage}{0.97\columnwidth}
\begin{lstlisting}[caption={Function Prompt Example}, label=lst:prompt_format]
def async_close(self, **kwargs: Any) -> bool:
    """
    `async_close()` must be called at the very end of any script that uses the asynchronous `opena` feature. This calls `async_join()` first and then closes the thread pool used for the asynchronous operations.

    Returns:
        status (bool): True on success
    """
\end{lstlisting}
\end{minipage}
\end{center}

After prefix prompts are extracted, we feed them into each of the collaboratively trained models (Section \ref{sec:colmethods}) to get the models' generation outputs, and then measure the amount of training data memorization in the generated outputs.


\paragraph{\textbf{Memorization Detection.}}
\label{sec:mem-detection}

We can detect memorized data by comparing the similarity between the outputs generated by the models and the individual participants' datasets.
The availability of the organizations' training dataset in our study allows us to easily make the comparison to check if there are duplications between the generated code snippets and the training datasets.
We adapt the memorization detection technique from Yang et al.~\cite{yang2024unveiling}, which employs the Simian clone detection tool\footnote{https://simian.quandarypeak.com/} to detect Type-1 clones between the generated code and the training code. A Type-1 clone, or exact clone, refers to identical segments of code (with minimum six lines as the default setting in Simian). If the model produces these exact replicas, it strongly suggests memorization. Therefore, we classify such a clone as an instance of memorization.

\paragraph{\textbf{Memorization Evaluation.}}
\label{sec:mem-ratio}

To better quantify the extent of training data memorization in the model-generated code, we introduce the \textit{Memorization Ratio}, which is defined in the following. Given a set of specific prompts, the code model generates a set of code. Simian is then used to detect \textit{x} distinct blocks of code that are identical to some blocks of code in a training dataset; these $x$ blocks of code are considered as memorized code. The Memorization Ratio is then calculated by summing the numbers of lines within all the blocks and then dividing by the total number of lines in all the generated code. Mathematically, this can be represented as:
\begin{equation}
\text{Mem. Ratio} = \frac{\sum_{i=1}^{x}\mbox{lines of code in memorized block$_i$}}{\sum\mbox{ lines of code in all generations}}
\end{equation}

\section{Datasets}
\label{sec:dataset}
This section presents some characteristics of the datasets we collected from different organizations (Section \ref{sec:data-construction}) and performs some preprocessing for the following evaluation.

\paragraph{\textbf{Collecting Organization's Codebase.}}
We utilized Google's BigQuery to collect all open-source licensed Python files from the GitHub database, resulting in a total of 27,128,930 files, amounting to 188.3 GB of data. Additionally, to identify the repositories on GitHub that belong to a certain organization, we manually identified some repositories' names that are very likely related to Google, Microsoft, and Facebook, and use them \edits{to} extract organization' codebase.
Table \ref{tab:orga_repos} shows sample names of the repositories collected for the organizations.
In total, we collected three Python codebase, one for each organization. There were 125,847 files (1018.93 MB) for Google, 33,560 files (703.29 MB) for Microsoft, and 5,207 files (38.58 MB) for Facebook.

\begin{table}[h]
\centering
\caption{Organizations' Repositories}
\label{tab:orga_repos}
\resizebox{\columnwidth}{!}{
\begin{tabular}{@{}l l p{0.7\linewidth}@{}}
\toprule
\textbf{Org.} & \textbf{\# of Repos} & \textbf{Sample Repos}  \\ \midrule
Google (G)       & 32    & google, google-research, google-deepmind, etc.  \\
Microsoft (M)   & 10    & microsoft, Azure, MicrosoftEdge, etc.  \\
Facebook (F)    & 5     & facebook, facebookresearch, fbsamples, etc.  \\
\bottomrule
\end{tabular}
}
\end{table}

\paragraph{\textbf{Preprocessing and Splitting.}}
As there can be duplicate files or low-quality code in the codebases that may affect model training,
we respectively preprocessed each dataset using methods employed in the training of the CodeParrot and PyCodeGPT models~\cite{cert2022ijcai}. These methods are based on heuristics proposed by OpenAI's Codex~\cite{chen2021evaluating} and have been further refined and enriched.
Sample filtering criteria used are as follows:
\begin{itemize}[leftmargin=2em,nosep]

    \item Removal of duplicate code files with method MinHash + LSH.
    \item Filtering out files with a fraction of alphanumeric characters less than 0.25.
    \item Removing files containing the phrase "auto-generated" or similar within the first five lines.
\end{itemize}

We then split each codebase into a training set and a validation set to facilitate model training.
Basic statistics of the resultant codebases are shown in Table \ref{tab:dataset_splits}.

\begin{table}[h]
\centering
\caption{Dataset Splits}
\label{tab:dataset_splits}
\small
\begin{tabular}{@{}llrr@{}}
\toprule
Dataset & Split       & Files Count  & Size       \\ \midrule
Google       & Training    & 53,545 & 501.77 MB \\
             & Validation  & 13,387 & 125.97 MB \\ 
Microsoft    & Training    & 15,251 & 327.47 MB \\
             & Validation  & 3,813  & 87.63 MB  \\ 
Facebook     & Training    & 2,148 & 19.34 MB \\
             & Validation  & 538  & 4.99 MB  \\              
             \bottomrule
\end{tabular}
\end{table}

\paragraph{\textbf{Cross-Org Codebase Characteristics.}}

As shown in Table \ref{table:dataset_per_mb}, we measured average metrics per megabyte, including lines of code (LOC), number of classes, number of functions, and number of docstrings across three datasets. Notably, there are discernible variations in these metrics among the datasets.

\begin{table}[h!]
\centering
\caption{Basic Metrics Per Megabyte}
\label{table:dataset_per_mb}
\small
\begin{tabular}{l r r r r}
\toprule
Dataset & LOC & Classes & Funcs & Docs \\
\hline
Google & 5,309.70 & 184.87 & 946.05 & 547.44 \\
Microsoft & 4,823.18 & 173.50 & 460.67 & 358.63 \\
Facebook & 7,588.45 & 248.58 & 1,435.30 & 557.21\\ 
\bottomrule
\end{tabular}
\footnotesize
\begin{flushleft}
\textbf{Note:} LOC - Lines of Code, Classes - Number of classes, Funcs - Number of functions, Docs - Number of docstrings.
\end{flushleft}
\end{table}

\begin{figure}[h]
\centering
\framebox[0.9\columnwidth]{
    \includegraphics[width=0.9\columnwidth]{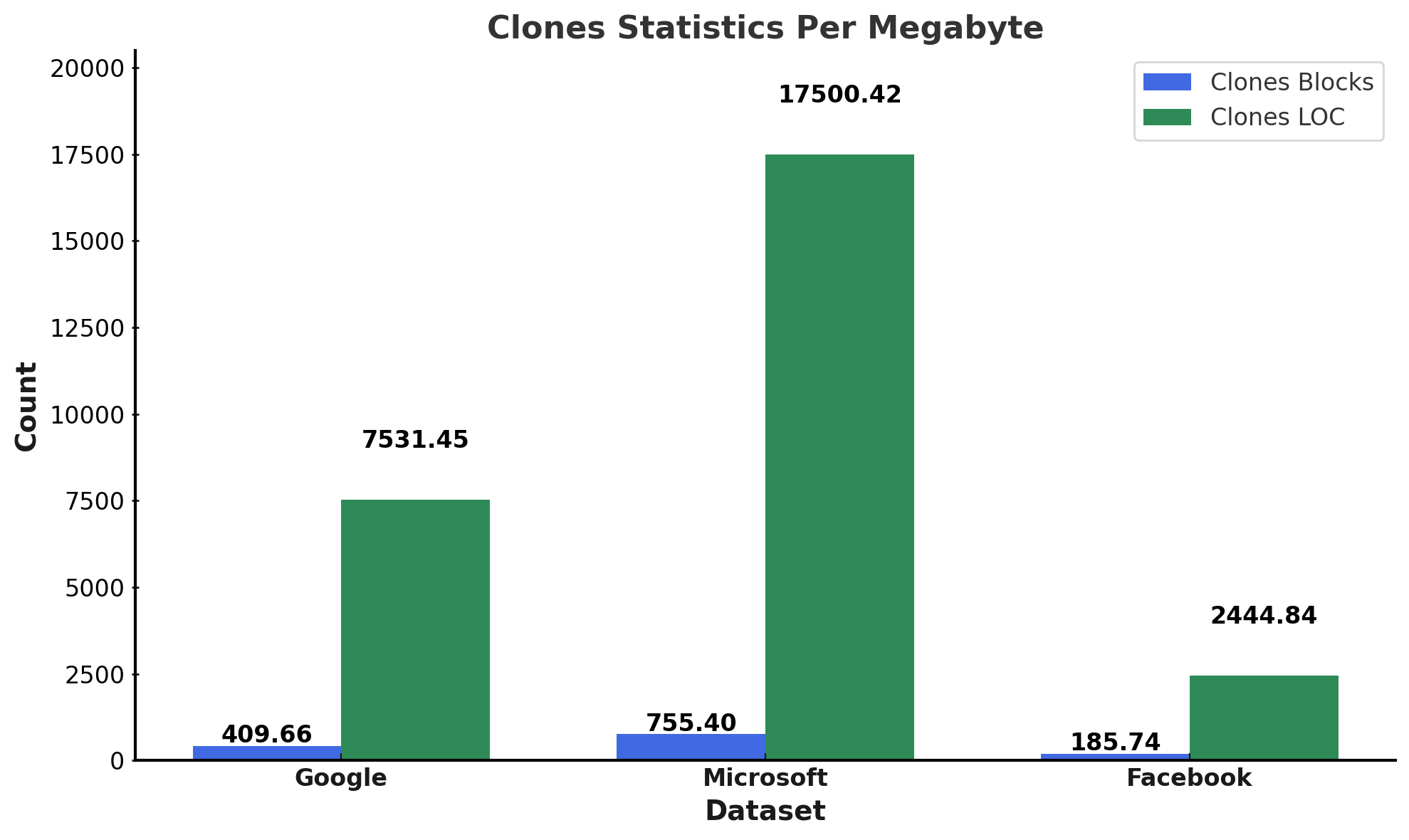} 
}
\caption{Clones Statistics Per Megabyte}
\label{fig:clones_stat}
\end{figure}

Internal clone detection was then performed with a threshold of minimum six lines to define a clone block for each organization's datasets. Figure \ref{fig:clones_stat} shows the clone statistics per megabyte for datasets from Google, Microsoft, and Facebook. The metrics analyzed include the average number of clone blocks and the lines of code (LOC) within these clone blocks. We examined these statistics to understand the extent of duplicated content within the datasets, as the findings by Yang et al.~\cite{yang2024unveiling} suggest that the occurrence of duplicate samples is correlated with an increased tendency for data memorization by code models.

The result reveals that the Microsoft dataset has the highest number of clone blocks (755.40) and clone LOC (17500.42) per megabyte, indicating a significant presence of duplicated content, which could lead to increased memorization during model training. In comparison, the Google and Facebook datasets have fewer clone blocks and LOC, indicating relatively lower duplication.

\section{Experiment Setup}
\label{sec:exp-setup}
This section describes the specific experimental settings implemented to provide answers to each research question.

\subsection{Base Model}
\label{sec:model_choice}

To minimize interference from existing training data, we chose \textit{GPT-2} as our base model because it was trained on the WebText dataset~\cite{radford2019language}, which includes substantial web data but not specifically GitHub code. This choice ensures that the Python dataset we collected from GitHub is relatively new to GPT-2's training data, reducing the potential memorization effect of the base model. Although there are other models that meet our requirements, we selected GPT-2 because it serves as the foundation for many widely used models, such as CodeParrot. 

\begin{table}[h!]
\centering
\caption{Comparison of Large Language Models}\label{tab:comparison_llms}
\resizebox{\columnwidth}{!}{%
 \begin{tabular}{l l r r}
\toprule
\textbf{Model} & \textbf{Training Data} & \textbf{Log(PPL)/Zlib} & \textbf{Release Date} \\
\hline
\multicolumn{4}{c}{\textbf{Basic LLMs (\char`\~
 125M Parameters)}} \\
\hline
GPT-2 & WebText (8M web documents) & 0.0020256 & Feb 18, 2019 \\
GPT-Neo-125M & The Pile(include code repos) & 0.0014813 & Apr 6, 2021 \\
CodeParrot-Small & Python code from GitHub & 0.0008022 & Nov 5, 2021 \\
PyCodeGPT & Python scripts from Github& 0.0008196 & Jan 4, 2023 \\
\hline
\multicolumn{4}{c}{\textbf{Advanced LLMs (\char`\~7/8B Parameters)}} \\
\hline
LLaMA-2-7B & Web, books, code data & 0.0006497 & Jul 18, 2023 \\
Mistral-7B-v0.1 & Mixed web and code data & 0.0007225 & Sep 20, 2023 \\
CodeLlama-7B-Python & High-quality Python code & 0.0003863 & Mar 14, 2024 \\
LLaMA-3-8B & Enhanced web and code data & 0.0006599 & Apr 17, 2024 \\
\bottomrule
\end{tabular}
}
\end{table}

To assess potential overlap between our collected Python codebase and LLMs' pre-training data, we employed a membership inference attack using the PPL-Zlib Ratio metric, which measures the ratio of log perplexity to zlib entropy~\cite{yang2024unveiling}. A lower ratio suggests that a code snippet was likely seen during pre-training~\cite{carlini2021extracting}. We sampled 10\% of our data and computed the average ratio for each LLM. As shown in Table \ref{tab:comparison_llms}, models specifically trained on Python code, such as CodeParrot, PyCodeGPT, and CodeLlama-Python, display lower ratios compared to similarly sized models. Advanced models like LLaMA-3 and Mistral, which also include code in their training, similarly show low ratios. Given the unavailability of real-world proprietary codebases and the risk of data leakage from GitHub repositories, GPT-2's higher PPL-Zlib Ratio, compared to models trained on Python code, suggests a lower likelihood of overlap with our dataset, making it a more suitable choice for our evaluation of collaborative training scenarios. 

\subsection{Collaborative Training}
\label{sec:training}
We conducted all collaborative training using the \textit{NVIDIA A100-PCIE-40GB GPU}, which features 40GB of high-bandwidth memory.

\paragraph{\textbf{Training Settings.}}

\begin{itemize}[nosep,leftmargin=1em]
\item For centralized learning (CL), we aggregated the three codebases into a single dataset and trained the model for 10 epochs.
Due to computational resource constraints, we set the training batch size to 2. For other hyperparameters, we followed the configurations used by CodeParrot\footnote{https://huggingface.co/codeparrot/codeparrot-small}. 
\item For federated learning (FL), we conducted a total of 10 rounds of training, with each client training on its own codebase for 1 epoch each during each round using the Flower federated learning framework.\footnote{https://flower.ai/docs/framework/tutorial-series-what-is-federated-learning.html} For the federated learning aggregation methods FedAvg and FedYogi, we utilized the default hyperparameters implemented in previous work~\cite{reddi2021adaptive} and maintained the same setup as centralized learning for each client's own training.
\item For incremental learning (IL), we considered all six distinct sequences for the three codebases: Facebook (F), Microsoft (M), and Google (G). The models were trained sequentially on each codebase for 10 epochs using the same hyperparameters as in the centralized setting.
\end{itemize}

\paragraph{\textbf{Trained Models.}}
We obtained nine collaborative models from collaborative training: one centralized model (\textit{Centralized\_FMG}), two federated learning models (\textit{Federated\_Avg\_FMG} and \textit{Federated\_Yogi\_FMG}), and six incremental learning models (\textit{Incremental\_{SEQUENCE}}). The SEQUENCE represents the training order of the datasets from Facebook (F), Microsoft (M), and Google (G) (either F2M2G, F2G2M, M2F2G, M2G2F, G2F2M, or G2M2F). Additionally, we trained three baseline models, one for each dataset (\textit{Facebook\_Only}, \textit{Microsoft\_Only} and \textit{Google\_Only}) for comparison with various collaborative models.

\subsection{Effectiveness Evaluation Settings for RQ1}


The combined evaluation dataset, derived from the unseen validation datasets of all participants, is used to calculate the Perplexity score. To assess correctness and utility, we estimated the pass@k metric in the \textit{EvalPlus} benchmark with $n\_samples=200$, following the settings of previous work~\cite{chen2021evaluating}, performing sampling with temperatures ranging from 0.1 to 1.0, and selecting the optimal value for each metric, as outlined in earlier research~\cite{cert2022ijcai}.

\subsection{Memorization Evaluation Settings for RQ2}

\paragraph{\textbf{Prompt Construction.}}
\label{sec:train_promp_construct}

We followed the prompt construction method from Section \ref{sec:data-extract}, extracting all functions and selecting signatures and docstrings of appropriate lengths as prompts. Given the large number of prompts obtained, we randomly sampled 10\% of the function prompts from each codebase for code generation. The outcomes of our prompt construction are presented in Table \ref{tab:function_counts}.


\begin{table}[h]
\centering
\caption{Function Prompts}
\label{tab:function_counts}
\small
\begin{tabular}{l r r r}
\toprule
\textbf{Dataset} & \textbf{Total Functions} & \textbf{Total Prompts} & \textbf{Sampled}\\
\hline
Google           & 475,256 & 187,900 & 18,790 \\
Microsoft        & 150,248 & 58,068 & 5,807 \\
Facebook         & 27,340  & 8,159  & 816 \\
\bottomrule
\end{tabular}
\end{table}

\paragraph{\textbf{Data Extraction and Memorization Evaluation.}}
To ensure efficient data extraction, we set the \textit{temperature} to 0.6 and the \textit{top-p (nucleus sampling)} to 0.6, following best practices outlined by Yu et al.~\cite{yu2023bag}, which assess various techniques for enhancing the training data extraction process from language models. Additionally, we configured the number of \textit{generations per prompt} to 5 and limited the \textit{maximum number of newly generated tokens} to 512. 
For the memorization evaluation, we followed the methods described in Section \ref{sec:mem-detection},
using the Simian tool with a default threshold of a minimum of 6 lines
 of code to report Type-1 clones, which are considered instances of memorization.


\subsection{Cross-Org Clone Memorization Evaluation Settings for RQ3}
\label{sec:setup_clones_eval}


\paragraph{\textbf{Collecting Cross-Organizational Clones.}}
To evaluate how such clones are memorized in collaborative models, we first identified the clones within the training datasets using the Simian tool, applying a default threshold of six consecutive lines to define a clone. Since these clones may not be complete functions and there are no function headers or docstrings, we need to construct prompts for the clones differently from Section \ref{sec:prompt-construction}: we chose to use the first half of a clone snippet as prefix prompts, and fed them to the models to generate the rest of the code.
For particularly long clones exceeding the 1024-token limit of GPT-2 after tokenization, we split them into smaller portions before creating the prefixes and suffixes.



First, we detected cross-organizational clones that are common across all three training data. However, due to the significant size disparity among these datasets---19.34 MB for Facebook, 327.47 MB for Microsoft, and 501.77 MB for Google, there were only 41 common clones. The limited number of prefix prompts for common clones led to inconclusive results. To enhance the evaluation of cross-organizational clones and obtain more robust and evaluable samples, we focused only on the Microsoft and Google datasets. These two datasets are larger, allowing for more clones. We identified 316 common clones between Microsoft and Google, encompassing a total of 7,536 lines (see Table 
\ref{tab:crosssilo_clones}). To manage extra-long clones, we divided them into smaller portions to ensure the tokenized lengths of the prefixes were under 512 tokens. This process resulted in 349 prefix prompts, providing a more substantial basis for evaluation.


\begin{table}[h]
\centering
\caption{Cross-silo Clones}
\label{tab:crosssilo_clones}
\small
\begin{tabular}{c c c c }
\toprule
\textbf{Dataset} & \textbf{Total Lns} & \textbf{Clone Blks} & \textbf{Clone Lns}  \\
\hline

Google & 12,040,783 & \multirow{2}{*}{316} &  \multirow{2}{*}{7536}  \\

\cline{1-2}

Microsoft & 6,501,197 & &  \\
\bottomrule
\end{tabular}
\footnotesize
\begin{flushleft}
\textbf{Note:} Total Lns: Total lines of code in the training data. Clone Blks: Num of clone blocks. Clone Lns: The total line counts of the clone blocks.
\end{flushleft}
\end{table}


\paragraph{\textbf{Model Training with Two Datasets Only.}}
To better evaluate the memorization of cross-organizational clone  from Microsoft (M) and Google (G) codebases, we trained additional models: \textit{Centralized\_MG}, \textit{Federated\_Avg\_MG}, \textit{Federated\_Yogi\_MG}, \textit{Incremental\_M2G}, and \textit{Incremental\_G2M}. These models were trained using only the two datasets with the same hyperparameter settings described in Section \ref{sec:training}.

\paragraph{\textbf{Clones Memorization Detection and Evaluation.}}
Finally, we applied the same methods as described in Section \ref{sec:mem-eval} to assess the memorization of cross-organizational clones, but using a detection threshold of three lines instead of six, excluding the other minimal three lines used as prefixes.

\section{Empirical Evaluation Results}

\subsection{Effectiveness of Collaborative Models}

The evaluation results for 
\textit{RQ1: 
What factors most significantly impact the effectiveness of collaborative training methods for code generation models?
} 
are shown in Table \ref{tab:ppl_passk}.


\begin{table}[h]
    \centering
    \caption{Perplexity Scores and Pass@k Results}
    \label{tab:ppl_passk}
    \resizebox{\columnwidth}{!}{%
    \begin{tabular}{l r r r r}
        \toprule
        \textbf{Model} & \textbf{Perplexity} & \textbf{Pass@1} & \textbf{Pass@10} & \textbf{Pass@100} \\
        \midrule
        \multicolumn{5}{c}{\textbf{Baseline}} \\
        \hline
        GPT-2 & 1084.78 & 0.0\% & 0.0\% & 0.0\% \\
        Facebook\_Only & 181.54 & 0.0\% & 0.0\% & 0.0\% \\
        Microsoft\_Only & 39.52 & 0.009\% & 0.090\% & 0.763\% \\
        Google\_Only & 5.36 & 0.213\% & 1.190\% & 3.281\% \\
        \hline
        \multicolumn{5}{c}{\textbf{Synchronous Collaborative Settings}} \\
        \hline
        Centralized\_FMG & 3.32 & 0.058\% & 0.530\% & 2.338\% \\
        Federated\_Avg\_FMG & 3.71 & 0.598\% & 2.212\% & 3.943\% \\
        Federated\_Yogi\_FMG & 4.02 & 0.506\% & 1.781\% & 3.770\% \\
        \hline
        \multicolumn{5}{c}{\textbf{Asynchronous Collaborative Settings}} \\
        \hline
        Incremental\_F2M2G & 5.30 & 0.546\% & 1.588\% & 3.800\% \\
        Incremental\_F2G2M & 26.34 & 0.003\% & 0.030\% & 0.305\% \\
        Incremental\_M2F2G & 5.36 & 0.521\% & 1.630\% & 3.165\% \\
        Incremental\_M2G2F & 37.45 & 0.012\% & 0.121\% & 1.068\% \\
        Incremental\_G2F2M & 25.42 & 0.0\% & 0.0\% & 0.0\% \\
        Incremental\_G2M2F & 53.38 & 0.006\% & 0.060\% & 0.458\% \\
        \bottomrule
    \end{tabular}
    }
\end{table}

\begin{table*}[h]
    \centering
    \caption{Memorization Evaluation Results for Different Training Datasets}
    \label{tab:combined-mem-evaluation}
    \resizebox{\textwidth}{!}{%
    \begin{tabular}{l r r r r | r r r r | r r r r}
        \toprule
        \textbf{Model} & \multicolumn{4}{c|}{\textbf{Google}} & \multicolumn{4}{c|}{\textbf{Microsoft}} & \multicolumn{4}{c}{\textbf{Facebook}} \\
        \cmidrule(lr){2-5} \cmidrule(lr){6-9} \cmidrule(lr){10-13}
        & \textbf{Lns of Gen.} & \textbf{Mem. Blks} & \textbf{Mem. Lns} & \textbf{Mem. Ratio} 
        & \textbf{Lns of Gen.} & \textbf{Mem. Blks} & \textbf{Mem. Lns} & \textbf{Mem. Ratio} 
        & \textbf{Lns of Gen.} & \textbf{Mem. Blks} & \textbf{Mem. Lns} & \textbf{Mem. Ratio} \\
        \midrule
        \multicolumn{13}{c}{\textbf{Synchronous Training Settings}} \\
        \hline
        Centralized\_FMG & 2,961,075 & 554 & 7,372 & 0.249\% & 994,977 & 4,732 & 63,215 & \textcolor{red}{6.353\%} & 170,951 & 4 & 40 & 0.023\% \\
        Federated\_Avg\_FMG & 2,954,757 & 723 & 7,799 & 0.263\% & 1,305,876 & 901 & 6,753 & 0.517\% & 168,833 & 11 & 112 & 0.066\% \\
        Federated\_Yogi\_FMG & 3,014,251 & 899 & 12,253 & 0.407\% & 1,492,648 & 8 & 56 & 0.004\% & 162,671 & 3 & 31 & 0.019\% \\
        \midrule
        \multicolumn{13}{c}{\textbf{Asynchronous Training Settings}} \\
        \hline
        Incremental\_F2M2G & 3,254,489 & 816 & 9,477 & \textcolor{red}{0.291\%} & 1,507,980 & 26 & 173 & 0.011\% & 173,845 & 4 & 37 & 0.021\% \\
        Incremental\_F2G2M & 2,900,802 & 98 & 877 & 0.030\% & 955,695 & 4,893 & 68,512 & \textcolor{red}{7.169\%} & 136,075 & 3 & 31 & 0.023\% \\
        Incremental\_M2F2G & 3,017,063 & 870 & 11,235 & \textcolor{red}{0.372\%} & 1,271,080 & 4 & 25 & 0.002\% & 162,598 & 4 & 40 & 0.025\% \\
        Incremental\_M2G2F & 2,463,402 & 63 & 576 & 0.023\% & 1,120,943 & 1 & 8 & 0.001\% & 120,339 & 23 & 188 & \textcolor{red}{0.156\%} \\
        Incremental\_G2F2M & 2,911,387 & 53 & 514 & 0.018\% & 967,170 & 5,762 & 78,632 & \textcolor{red}{8.130\%} & 136,737 & 2 & 23 & 0.017\% \\
        Incremental\_G2M2F & 1,932,255 & 62 & 528 & 0.027\% & 714,779 & 1 & 6 & 0.001\% & 96,647 & 14 & 146 & \textcolor{red}{0.151\%} \\
        \bottomrule
    \end{tabular}
    }
\end{table*}

\paragraph{\textbf{Baseline Models.}}
The \textit{GPT-2} base model, not specifically trained on the provided datasets, showed poor performance with a high perplexity score and 0\% pass rates across all k values in the Pass@k metric. Among the individual dataset models, the \textit{Google\_Only} model exhibited the best performance, underscoring the importance of a larger dataset size for better model effectiveness.

\paragraph{\textbf{Synchronous Collaborative Settings.}} 
In synchronous collaborative training settings, the \textit{Centralized\_FMG} model demonstrated the best performance in next token prediction ability with the lowest perplexity score of 3.32. The two federated models, \textit{Federated\_Avg\_FMG} and \textit{Federated\_Yogi\_FMG}, also achieved comparable perplexity scores of 3.71 and 4.02, respectively. Overall, both centralized and federated models outperformed the incremental models in the perplexity metric. For the pass@k metric, federated learning models, particularly \textit{Federated\_Avg\_FMG} and \textit{Federated\_Yogi\_FMG}, surprisingly surpassed the \textit{Centralized\_FMG} model. This indicates that federated learning approaches can achieve effectiveness comparable to centralized training while keeping training data private.

\paragraph{\textbf{Asynchronous Collaborative Settings.}}
For asynchronous collaborative training settings, the effectiveness varied significantly based on the order in which datasets were used. The \textit{Incremental\_F2M2G} model performed the best among incremental models, suggesting that starting with smaller datasets and sequentially adding larger ones might be beneficial. 

\begin{summarybox}{Summary of Findings for RQ1}
Our evaluation underscores the importance of dataset size, diversity, and the order of data introduction in collaborative training. Federated learning emerged as a promising method, balancing privacy and performance better. However, the variability in effectiveness for incremental learning models highlights the need for careful planning and strategy when introducing datasets sequentially.
\end{summarybox}

\subsection{Memorization in Collaborative Models}

The evaluation results for \textit{RQ2: To what extent is data from different participants memorized in various collaborative training settings?}
are presented in Table \ref{tab:combined-mem-evaluation} and Table \ref{tab:mem-evaluation-scores-summed}.

\begin{table}[h]
    \centering
    \caption{Summed Memorization Results Across All Datasets}
    \label{tab:mem-evaluation-scores-summed}
    \resizebox{\columnwidth}{!}{%
    \begin{tabular}{l l r r r r}
        \toprule
        \textbf{Rank} & \textbf{Model} & \textbf{Lns of Gen.} & \textbf{Mem. Blks} & \textbf{Mem. Lns} & \textbf{Mem. Ratio} \\
        \midrule
        1 & Incremental\_G2F2M & 4,015,294 & 5,817 & 79,169 & 1.971\% \\
        2 & Incremental\_F2G2M & 3,992,572 & 4,994 & 69,320 & 1.736\% \\
        3 & Centralized\_FMG & 4,127,003 & 5,290 & 70,627 & 1.711\% \\
        4 & Federated\_Avg\_FMG & 4,439,466 & 1,635 & 14,664 & 0.330\% \\
        5 & Federated\_Yogi\_FMG & 4,679,570 & 910 & 12,340 & 0.264\% \\
        6 & Incremental\_M2F2G & 4,450,735 & 878 & 11,260 & 0.253\% \\
        7 & Incremental\_F2M2G & 4,936,314 & 846 & 9,687 & 0.196\% \\
        8 & Incremental\_G2M2F & 2,743,671 & 77 & 680 & 0.025\% \\
        9 & Incremental\_M2G2F & 3,704,684 & 87 & 772 & 0.021\% \\
        \bottomrule
    \end{tabular}
    }
\end{table}
\vspace{-0.5em}

\paragraph{\textbf{From a Dataset Perspective.}}
The Microsoft training data shows higher memorization compared to the other two datasets across different collaborative models. For example, in the \textit{Centralized\_FMG} model and the two incremental models ending with Microsoft datasets (\textit{Incremental\_F2G2M} and \textit{Incremental\_G2F2M}), the memorization ratios are 6.353\%, 7.169\%, and 8.130\%, respectively. This can be attributed to intrinsic differences among the datasets. As illustrated in Table \ref{table:dataset_per_mb} and Figure \ref{fig:clones_stat}, these datasets vary significantly in terms of internal duplicates, average file size, and the number of docstrings and functions. The high number of internal duplicates in the Microsoft dataset leads to increased memorization ratios across models. This aligns with the findings of Yang et al.~\cite{yang2024unveiling}, which indicate that frequently occurring code snippets in the training data are more likely to be memorized.

\paragraph{\textbf{From a Model Perspective.}}
The ranked memorization ratios in Table \ref{tab:mem-evaluation-scores-summed} indicate that the models \textit{Incremental\_G2F2M}, \textit{Incremental\_F2G2M}, and \textit{Centralized\_FMG} exhibit the highest overall memorization ratios, significantly higher than others. By examining Table \ref{tab:combined-mem-evaluation}, it becomes clear that this high memorization is largely due to their substantial retention of the Microsoft dataset, which contains a higher level of internal duplicates. Furthermore, when examining other incremental learning settings, it becomes evident that all models exhibit the highest memorization ratio for the last dataset they trained on. This trend is concerning for collaboration, as it suggests that the final dataset in the training sequence is memorized at a disproportionately higher ratio, increasing the risk for the last participants. Such a pattern raises significant concerns for participants considering the use of incremental learning settings, as the final participants might face greater risks of data leakage and privacy issues.

\vspace{-1em}
Notably, our experiments reveal that both federated learning methods, \textit{FedAvg}, which aggregates weights from different participants, and \textit{Yogi}, which adaptively adjusts the learning rate for non-IID datasets, exhibit relatively low levels of memorization across training datasets. This highlights federated learning as a promising approach for collaborative training, as it protects privacy better by keeping training data unseen during the training phase and maintains low memorization ratios during inference, all while achieving performance comparable to centralized models.

\begin{summarybox}{Summary of Findings for RQ2}
Our evaluation reveals that datasets with a higher number of internal duplicates exhibit greater memorization in collaborative training. Centralized models demonstrate relatively high memorization ratios, whereas incremental learning settings display unstable memorization ratios, heavily influenced by their training sequence, with the last dataset in the sequence being memorized at a disproportionately higher ratio. Federated learning methods, such as \textit{FedAvg} and \textit{Yogi}, maintain relatively low level of memorization for training data, highlighting their promise for collaborative training.
\end{summarybox}

\subsection{Cross-Org Clones Memorization Evaluation}

The evaluation results for \textit{RQ3: How are cross-organizational clones memorized in collaborative models?}
are presented in Table \ref{tab:mem-evaluation-scores}. 
\vspace{-0.5em}

\begin{table}[h]
    \centering
    \caption{Cross-Org Clones Memorization Evaluation Results}
    \label{tab:mem-evaluation-scores}
    \resizebox{\columnwidth}{!}{%
    \begin{tabular}{l r r r r}
        \toprule
        \textbf{Model} & \textbf{Lns of Gen.} & \textbf{Mem. Blks} & \textbf{Mem. Lns} & \textbf{Mem. Ratio} \\
        \midrule
        \multicolumn{5}{c}{\textbf{Synchronous Training Settings}} \\
        \hline
        Centralized\_MG & 46,173 & 55 & 261 & 0.565\% \\
        Federated\_Avg\_MG & 43,976 & 51 & 243 & 0.552\% \\
        Federated\_Yogi\_MG & 43,950 & 44 & 211 & 0.480\% \\
        \midrule
        \multicolumn{5}{c}{\textbf{Asynchronous Training Settings}} \\
        \hline
        Incremental\_G2M & 33,265 & 16 & 64 & 0.192\% \\
        Incremental\_M2G & 38,618 & 24 & 95 & 0.246\% \\
        \bottomrule
    \end{tabular}
    }

    \footnotesize
\begin{flushleft}
\textbf{Note:} Lns of Gen: Total lines generated. Mem Blks: Num of memorized blocks. Mem Lns: Total lines of memorized blocks. Mem. Ratio: ratio of Mem. Lns to Lns of Gen.
\end{flushleft}
\end{table}

Based on our observations, it is evident that in \textit{Synchronous Collaborative Training} settings, which include both Centralized Training and Federated Learning, models tend to exhibit higher memorization ratios for cross-organizational clones. This can be attributed to the repetitive learning of these clones during each weight update across multiple datasets. Specifically, the \textit{Centralized} model shows a memorization ratio of 0.565\%, while the \textit{Federated\_Avg\_MG} and \textit{Federated\_Yogi\_MG} models demonstrate memorization ratios of 0.552\% and 0.480\%, respectively. In contrast,  cross-organizational clones are memorized at a relatively lower ratio in incremental learning settings. We believe this is due to catastrophic forgetting~\cite{shi2021overcoming}, where a model trained sequentially on different tasks or datasets tends to overwrite the knowledge gained from previous tasks with new information from the current task. Consequently, incremental learning models exhibit a lower memorization ratio than their synchronous counterparts, with the \textit{Incremental\_G2M} model having the lowest memorization ratio at 0.192\%, followed by the \textit{Incremental\_M2G} model at 0.246\%.

\vspace{-1.0em}
Our findings indicate that memorization of cross-organizational clones is relatively higher in centralized and federated settings. Notably, in the context of federated learning, participants are restricted to processing their own datasets, making it challenging to reduce cross-organizational clones. Redundant training on these clones not only wastes valuable computing resources but also leads to unbalanced feature learning and increases the risk of memorization. This highlights a critical need in federated learning to deduplicate clones across distributed datasets. Addressing this issue is essential to 
ensure the trustworthiness of collaborative models.

\begin{summarybox}{Summary of Findings for RQ3}
Our evaluation of cross-organizational clones in collaborative training settings revealed that synchronous methods, such as centralized training and federated learning, exhibit higher memorization ratios of cross-organizational clones than asynchronous methods like incremental learning. This underscores the need for effective preprocessing strategies in collaborative training scenarios to handle cross-organizational clones, ensure balanced feature learning, optimize computational resources, and mitigate memorization, especially when datasets are decentralized and access to them is restricted.
\end{summarybox}

\section{Discussion}
\label{sec:discuss}

\subsection{Suggestions to Practitioners}
For practitioners, it is crucial to focus on the size, diversity, and internal duplicates of datasets. A diverse, well-preprocessed dataset can significantly enhance performance and reduce memorization risks in collaborative training settings. Federated learning methods like FedAvg and FedYogi strike a good balance between performance and privacy preservation, maintaining low memorization ratios while achieving performance comparable to centralized training, making them ideal for scenarios prioritizing data privacy. In incremental learning, the sequence of dataset introduction must be carefully planned, as the final dataset in the sequence is more prone to memorization. 

Notably, practitioners should be vigilant about data leakage risks during inference. Even privacy-preserving methods like federated learning, which ensure the training data remains unseen and maintain a relatively low memorization ratio, can still produce verbatim code snippets from hidden training data, potentially violating data privacy or copyright. Therefore, implementing additional techniques, such as differential privacy~\cite{wei2020federated}, should be considered to mitigate these risks.


\subsection{Suggestions to Researchers}

For future research directions, it is imperative to explore integrating these settings with additional privacy protection techniques, such as random perturbation~\cite{li2023robin} and differential privacy~\cite{latif2020introducing}. While these techniques offer more privacy-preserving approaches, their potential impact on performance must be carefully considered. Evaluating the combination of these techniques can help strike a balance between the code generation model’s performance and the memorization of participants’ training data. Furthermore, it is crucial to explore advanced preprocessing techniques applicable in distributed training environments. Reducing clones across different datasets can save computing resources and prevent repetitive training on duplicate content, thereby enhancing overall model efficiency. 

Moreover, investigating collaboration during the inference phase, in addition to our study's focus on the training phase, would be beneficial for exploring more real-world collaboration options. Techniques such as ensemble learning~\cite{zou2021multi}, which combines knowledge from multiple models, or the ChatDev model~\cite{qian2023communicative}, which leverages natural language communication among agents to streamline collaborative development, could offer valuable insights.


\subsection{Threats to Validity}

\paragraph{\textbf{Threats to Internal Validity.}}
Several factors may threaten the internal validity of our study. Differences in dataset size, quality, and diversity from Facebook, Microsoft, and Google could introduce biases. Hyperparameter choices may have influenced performance, and while we used recommended settings, optimal values could vary. The implementation details of federated and incremental learning algorithms, such as aggregation strategy and data order, might have affected the results. We adhered to best practices to minimize these threats but acknowledge potential variations. Another potential threat to internal validity is our choice of GPT-2 as the base model, which is less powerful and advanced compared to more recent models like LLaMA-3 or Mistral. However, the same experimental setup was applied across all collaborative training scenarios, ensuring consistent improvements or degradations, regardless of the base model. For example, as Yang et al.~\cite{yang2024unveiling} demonstrated, more powerful models tend to memorize more training data. Therefore, the observed patterns in different collaborative scenarios should still hold true. Moreover, as discussed in Section \ref{sec:model_choice}, GPT-2 was selected due to the unavailability of real-world proprietary codebases and to mitigate the risk of data leakage from our collected GitHub codebase. While our preliminary work lays the groundwork for understanding collaborative training scenarios, it also highlights the need for industry and academia to collaborate. With access to large-scale private codebases, we could leverage more advanced models like LLaMA-3 to conduct a more thorough evaluation of collaborative training in large code models.

\paragraph{\textbf{Threats to Construct Validity.}}
A potential threat to construct validity is whether the metrics (perplexity, pass@k, and memorization ratio) and prompts (function signatures and docstrings) are appropriate and sufficient for measuring model performance and memorization. To mitigate this threat, we use a combination of metrics to comprehensively evaluate syntactic accuracy and the practical utility of generated code. Additionally, the prompts were constructed from realistic targeted attack scenarios to ensure relevance to potential real-world usages. This multi-faceted approach ensures that our evaluation reflects the constructs of interest, enhancing the validity of our findings. Another threat to construct validity is the potential impact of deduplication methods on the quality of participants' training data. Different models use various deduplication approaches, such as SHA256 hashing for exact file deduplication (e.g., CodeGen~\cite{nijkamp2022codegen}, PolyCode~\cite{xu2022systematic}), Levenshtein distance (PaLM Coder~\cite{chowdhery2023palm}), and MinHash + LSH for near-duplication applied in CodeParrot and StarCoder~\cite{li2023starcoder}. To mitigate this threat, we used the same near-deduplication methods as CodeParrot to ensure consistency and minimizing bias. Future research could explore substring deduplication\footnote{https://huggingface.co/blog/dedup}, balancing diversity and redundancy.

\paragraph{\textbf{Threats to External Validity.}}
The generalizability of our findings may be limited by focusing on datasets from three major technology companies, which may not represent other domains. Different model architectures might behave differently under collaborative training methods. Our performance and memorization metrics might not capture all aspects of model quality. The collaborative training scenarios we investigated may not cover all real-world cases. To address these threats, we used diverse settings to enhance the validity of our findings.\footnote{For memorization detection, we also tested thresholds of 4 and 8, which showed high correlation with the default threshold of 6. Consequently, we use the default threshold. Results for thresholds 4 and 8 are available in the replication package, under the /appendices directory.}
 Besides, there are other collaborative learning methods beyond the three kinds of collaborative {training} methods studied in this paper, e.g., voting-based ensemble learning \cite{opitz1999popular}, which focus on collaboration that happen during the inference/generation phase, instead of the training phase. They may have different collaboration mechanisms and allow models trained by individual participants to have very different structures. It can be interesting future work to investigate more kinds of collaborative models. Another threat to external validity is that the code generation evaluation benchmark we used may not fully represent real-world code generation scenarios. Our training codebase is sourced publicly from GitHub, which may not be comparable to the private and more complex source code used in real-world commercial software. To address this limitation, we focused on function generation tasks rather than more intricate real-world scenarios. To evaluate the effectiveness of different collaborative settings, we selected EvalPlus~\cite{liu2024your}, a modern and advanced peer-reviewed benchmark. EvalPlus extends the popular HumanEval~\cite{chen2021evaluating} benchmark with 80 times more test cases, making it a suitable choice for our study. For future work, we recommend collaboration with academia and industry partners to conduct larger-scale collaborative training experiments with more comprehensive and complex proprietary codebase and advanced base models. This would allow evaluation on more complex benchmarks like BigCodeBench\cite{zhuo2024bigcodebench}, which was released in June 2024 and is currently under peer review as of the writing of our study.

\section{Related Work}

\subsection{Memorization in Large Code Models}
Memorization in large code models is a significant concern in software engineering~\cite{yang2024robustness}. Research ~\cite{ciniselli2022extent, rabin2023memorization} indicates that code recommendation models often memorize numerous clones from their training data. Additionally, recent studies have revealed that sensitive information can potentially be leaked or extracted by large language models for code (LLM4Code) ~\cite{huang2024your,Al-Kaswan2024memorization,niu2023codexleaks,yang2024unveiling}. For instance, Niu et al.~\cite{niu2023codexleaks} designed prompts likely to induce privacy information from GitHub Copilot, discovering that approximately 8\% of these prompts resulted in privacy leaks. Yang et al.~\cite{yang2024unveiling} examined how large-scale datasets and advanced architectures lead to models inadvertently memorizing and reproducing code snippets verbatim, posing security and privacy risks. Their study categorizes memorized content, identifies exacerbating factors, and offers mitigation strategies. Al-Kaswan et al.~\cite{Al-Kaswan2024memorization} compared memorization in code-specific large language models (LLMs) to those trained on natural language, highlighting the susceptibility of code LLMs to data extraction attacks. Their findings emphasize the need for further exploration into memorization to develop effective safeguards against data leakage. Huang et al.~\cite{huang2024your} found multiple instances of credentials generated by neural code completion tools, including two that successfully authenticated real online service APIs.

Our research extends these investigations by exploring memorization challenges in models trained on distributed datasets under complex collaborative training settings. We aim to deepen the understanding of memorization concerning participants' training data and cross-organizational clones, fostering the development of secure, privacy-conscious collaborative code generation models.

\subsection{Federated Learning for SE}
Federated Learning (FL) in Software Engineering field has primarily focused on tasks such as defect prediction, code clone detection, and code summarization, which offer deterministic outputs. For instance, Yang et al.'s ALMITY~\cite{yang2024federated} and Kumar et al.'s FedLLM~\cite{kumar2024codesummarization} enhance model performance on skewed data distributions while ensuring data privacy. Yamamoto et al.~\cite{yamamoto2023towards} explored FL for Cross Project Defect Prediction (CPDP), preserving data privacy while maintaining competitive performance. Zhang et al.~\cite{zhang2024vulnerability} introduced a federated learning-based framework for vulnerability detection, and Alawdi et al.~\cite{alawadi2024fedcsd} developed FedCSD for code-smell detection, enabling collaborative training while safeguarding data privacy. Our research diverges by applying federated learning to the generative task of code generation. Unlike defect prediction or bug detection tasks, code generation involves creating executable code based on diverse specifications, presenting greater unpredictability and a higher risk of leaking participants' training data. By comparing model effectiveness and training data memorization with centralized and incremental learning settings, our study uncovers both the potential and the memorization risks of federated learning for the code generation task in a collaborative training scenario.

\section{Conclusion}
Our research highlights that the size and diversity of datasets are critical for the success of various collaborative training approaches for code generation. Specifically, federated learning models showed performance on par with centralized training while maintaining a relatively low memorization ratio, making them ideal for privacy-preserving training. Conversely, centralized training exhibited higher memorization ratios, especially for codebase with numerous internal duplicates. Furthermore, the sequence of dataset introduction significantly influenced the effectiveness and memorization patterns in incremental learning. Additionally, cross-organizational clone memorization is more prevalent in centralized and federated learning settings, underscoring the need for specialized preprocessing for decentralized datasets. Importantly, our study emphasizes that data leakage risks persist during the inference phase, even when strategies are employed to ensure the training data remains unseen during the collaborative training. We offer practical and insightful recommendations for both practitioners and researchers to enhance privacy- and copyright-preserving capabilities and facilitate cross-organizational collaboration for code generation. By doing so, we can better leverage the untapped value of segregated code datasets, thereby driving advancements in code generation models.

\vspace{1em}
\begin{mdframed}[backgroundcolor=gray!10, linecolor=black, linewidth=0.5pt, innerleftmargin=5pt, innerrightmargin=5pt, innertopmargin=5pt, innerbottommargin=5pt]
\textbf{Replication Package}: To support reproducibility, verification, and further research, we are providing our scripts, datasets, and prompts at this URL: \url{https://osf.io/7486g/?view_only=39b6cbb0c9d54439aabff52ad4aa827b}
\end{mdframed}

\begin{acks}
We acknowledge The Simian Similarity Analyzer, developed by Simon Harris and Quandary Peak Research, which was used for our memorization detection experiments. We extend our gratitude to Zhou Yang for his valuable guidance during the memorization detection experiments. The Simian Similarity Analyzer is © 2023–2024 Quandary Peak Research. This research is supported by the Ministry of Education, Singapore under its Academic Research Fund Tier 3 (Award ID: MOET32020-0004). Any opinions, findings and conclusions or recommendations expressed in this material are those of the author(s) and do not reflect the views of the Ministry of Education, Singapore.
\end{acks}

\balance
\printbibliography

\end{document}